\documentclass[twocolumn,nature]{revtex4}
\usepackage{epsfig}
\usepackage{amsmath} 
\usepackage{graphicx} 
\usepackage{verbatim} 
\usepackage{color} 
\usepackage{hyperref} 

\begin{document}

\title{Thermodynamics of Deposition Flux dependent Intrinsic Film Stress}

\author{Amirmehdi Saedi}
\affiliation{Huygens-Kamerlingh Onnes Laboratory, Leiden University, Niels Bohrweg 2, 2333 CA Leiden, The Netherlands\\Current address: ARCNL, Science Park 102, 1098 XG Amsterdam, The Netherlands}
\author{Marcel J. Rost}
\email{rost@physics.leidenuniv.nl}
\affiliation{Huygens-Kamerlingh Onnes Laboratory, Leiden University, Niels Bohrweg 2, 2333 CA Leiden, The Netherlands}

\date{\today}

\begin{abstract}
Vapor deposition on polycrystalline films can lead to extremely high levels of compressive stress, exceeding even the yield strength of the films. A significant part of this stress has a reversible nature: it disappears when the deposition is stopped and re-emerges upon resumption. Although the debate on the underlying mechanism still continues, insertion of atoms into grain boundaries seems to be the most likely one. However, the required driving force has not been identified. To address the problem we analyze, here, the entire film system using thermodynamic arguments. We find that the observed, tremendous stress levels can be explained by the flux induced entropic effects in the extremely dilute adatom gas on the surface. Our analysis justifies any adatom incorporation model, as it delivers the underlying thermodynamic driving force. Counterintuitively, we also show that the stress levels `decrease', if the barrier(s) for adatoms to reach the grain boundaries are `decreased'!
\end{abstract}





\maketitle

\section{Introduction}
\begin{figure}[h!tb]
\begin{center}
\epsfig{file=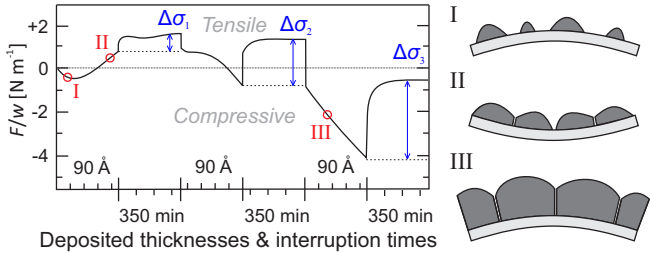,width=8.6cm} \caption{{\bf Stress evolution during VW-type film growth} for copper deposition with a flux of 0.1 \AA\ s$\mathrm{^{-1}}$ onto silicon oxide at room temperature. It consists of three main stages: nucleation (I), coalescence (II), and thickening (III). The deposition was interrupted 3 times for 350 minutes. The reversible stress jumps are indicated by $\Delta \sigma_{\mathrm{i}}$ (reproduced from \cite{Shu96}).}
\label{fig1}
\end{center}
\end{figure}

During the growth of a polycrystalline film on a substrate, the film usually develops a significant amount of internal stress. If the film temperature is high enough to reach Volmer-Weber (VW) type growth conditions \cite{Flo02,Spa00}, the film stress during deposition follows a compressive-tensile-compressive (CTC) evolution, as is indicated with stages I, II, and III in Fig. \ref{fig1}.
During stage I, the nucleated islands develop a compressive stress due to surface tension effects \cite{Lau81,Cam00}. Stage II occurs during film closure when the 3D islands coalesce and form grain boundaries (GBs). At this stage, the film free energy can be further lowered by `GB zipping', which in turn delivers tensile stress \cite{Hof76,Fre01}. Without sufficient mobility (low temperature or high deposition flux) the film remains tensile upon further growth. In contrast, a maximum tensile stress develops for VW-type growth of high mobility materials, which occurrence coincides approximately with the moment the film closes. From this moment on, the stress turns once again towards compressive values (stage III) \cite{Abe90}. Surprisingly, a significant part of the compressive stress has a reversible nature: upon interrupting the deposition flux (see Fig. \ref{fig1}), the film stress jumps to less compressive values and the original compressive stress state before interruption is almost fully restored when the flux is switched on again. These stress jumps can be as large as $\sim$150 MPa and the time constant of the stress variation upon resuming the deposition is in the order of 20 seconds \cite{Shu96,Fri04}!

In the last 20 years several mechanisms have been proposed aiming to explain the observed effects: 1) pre-coalescence surface tension continuation combined with ongoing grain growth \cite{abe79}, 2) surface roughness development during deposition combined with step-step-interactions \cite{Spa00}, 3) adatom insertion into GBs \cite{Cha02,Tel07,Pao07}, 4) interaction of adatoms with surface and each other \cite{Fri04}, 5) `inside bundling - outside grooving' of GBs  \cite{Gon13}, 6) depth changes in the GB grooves \cite{Yu14}, ... Whereas several of these mechanisms rely on kinetically limited processes, the GB adatom insertion model suggests that the compressive stress is generated via adatoms that are forced into the GBs by the enhanced chemical potential (CP) of the surface that is set up by the deposition flux. By switching off this flux, the CP should drop, which should lead to an outflow of the excess atoms from the GBs and, thereby, to a relaxation of the compressive stress \cite{Cha02}. Recent experiments confirmed that GBs are prerequisite for the existence of the reversible stress jumps \cite{Lei09}. However, more questions arise, as the time constant of the stress relaxation upon interruption seems to be temperature independent \cite{Lei10}. On the other hand, surface stress effects \cite{Fri04} are expected to be too low in magnitude \cite{Pao06} to explain the reversible stress jumps. While the discussion on the mechanism(s) still continues, at a more fundamental level, the underlying driving force behind the effect has never been addressed. In this paper we derive the magnitudes and the changes of the CP on the surface next to the position of the GBs and show that this indeed forms the driving force for any adatom insertion model.

From a thermodynamic point of view, the most fundamental question has never been addressed, probably due to conceptual difficulties in calculating the CP of the surface during the growth: `how can a flux (change) as low as $\sim$0.1 monolayer per second [ML\ $\mathrm{s^{-1}}$] lead to stress jumps as high as $\sim$150 MPa?' Our study focuses exactly on this question and we show not only that these low fluxes can generate such huge stresses, but also that the driving force for the stress jumps is decreased, if it is easier for the adatoms to diffuse towards and into the GBs.

\section{Results}
\subsection{Basic Thermodynamic Description}
To derive our model, it is important to realize that the film is under growth conditions and therefore naturally not in equilibrium. However, as long as the growth conditions do not change, it can be treated in steady-state, like the famous `Growth-Wulff construction' \cite{Sek04}. The enhanced surface CP with respect to equilibrium sets up an adatom current to steps, which finally leads to the film growth. This also means that the CP on the surface varies locally and that positions connected to each other will try to balance their difference. If atom transport is sufficiently active on the time scale of consideration, one can approximate adjacent positions to be in equilibrium. Therefore, for constant `small' deposition fluxes and the absence of kinetic limitations, thermodynamic equilibrium can be assumed between the positions on the surface immediately next to grain boundaries (s/GB), the grain boundaries (GB), and the grain interiors (g). This assumption is further underpinned by the small number of total additional atom that have to be incorporated in the GBs. For the surface we solve rate equations to determine the CP immediately next to the GBs and we further treat this position, the GBs, and the grain interior to be in equilibrium. Thermodynamic equilibrium certainly does not hold for the transition between the flux `on' and `off' state, but is justified a few tens of seconds after the flux change (see above). Therefore, at constant or zero flux, a change of the CP of the surface next to the GBs, will finally change the CP of the GBs with the same amount, which in turn will change the CP of the grains:
\begin{equation}
\label{equ1}
\Delta {{\mu }_{\mathrm{s/GB}}}\approx \Delta {{\mu }_{\mathrm{GB}}}\approx \Delta {{\mu }_{\mathrm{g}}}\Rightarrow \Delta {{\mu }_{\mathrm{s/GB}}}\approx \Delta {{\mu }_{\mathrm{g}}}
\end{equation}
This core equation enables us to bypass the determination of the CP of the GBs, as well as the absolute CP values on the surface and within the grains.

\subsection{Chemical Potential of the Surface}
The free energy of a surface depends on the formation and interaction energies of a myriad of surface features such as terraces, steps, kinks, step adatoms, adatoms,... As the surface morphology evolves during deposition, the population of these features changes accordingly. However, it is known that the reversible, compressive stress can develop within seconds after starting the flux even with rates as low as 0.1 ML\ $\mathrm{s^{-1}}$. Obviously, the population of point-like features, like adatoms, step-adatoms, and kinks, can (and will) change abruptly upon the arrival of flux on the surface, but extended surface features, like terraces and step edges, do not change significantly within such short time scales, as they consist of a large number of atoms \cite{Gie99-2}. For example, the surface roughness is directly linked to changes in the appearance, distribution, and amount of steps and terraces. The CTC behavior is usually observed under step-flow growth mode conditions, where changes in the surface roughness are known to happen very slowly \cite{Ros07, Mic04}. This means that the gradual increase in surface roughness (and hence the extended surface features) during deposition will only have a long time effect on the surface CP via the Gibbs-Thomson relation. Since the reversible stress jumps occur in a matter of seconds, we safely can ignore these long term changes in our analysis. Moreover, in contrast to the adatoms that live in a 2D-space on the terraces, the step-adatoms and step-kinks are confined to the 1D space on the step edges. This causes the rate, at which the step-adatoms and kinks meet and annihilate each other, to be significantly higher than the adatoms on the terraces. As a result, the increase in adatom population on the terraces, upon starting the deposition, is orders of magnitude higher than that of kinks or step-adatoms \cite{Zha90}. The conclusion is that the surface CP change that is responsible for the almost instantaneous stress jumps, is mainly dominated by a change in the adatom density. Consequently, we can neglect all other contributions, as they would lead only to higher order correction terms in determining the surface CP variations:
\begin{equation}
\label{equ2}
\begin{split}
& \Delta {{\mu }_{\mathrm{s}}} =\Delta {{\mu }_{\mathrm{adatom}}}+O(\Delta {{\mu }_{\mathrm{step\,\,adatom}}},\,\,\Delta {{\mu }_{\mathrm{kink}}},\,\,\Delta {{\mu }_{\mathrm{step}}},..)\\
& \approx\Delta \left[ \frac{\partial {{U}_{\mathrm{adatom}}}}{\partial N}+\frac{\partial {{U}_{\mathrm{adatom\,\,int.}}}}{\partial N}-T\frac{\partial {{S}_{\mathrm{adatom}}}}{\partial N} \right]
\end{split}
\end{equation}
The first term in the brackets, $\partial {{U}_{\mathrm{adatom}}}/\partial N$, accounts for the surface temperature dependent change in average energy (potential and kinetic) of individual adatoms. Given by the radiation of the evaporator and the kinetic energy of the arriving atoms, the increase in film temperature is less than 10 K for Cu, Ag, and Au \cite{Abe80}, which corresponds to $\sim$2.6 meV per film atom according to the classical Dulong-Petit limit of the heat capacity in solids. As these materials all have an excellent thermal conductivity, the surface and the bulk temperatures are virtually identical. Since we finally have to compare only the `CP variation' of the surface and the grains, we can safely neglect this term, as we would have to add the same value to both sides of Eq. \ref{equ1}.

The second term, $\partial {{U}_{\mathrm{adatom\,\,int.}}}/\partial N$, corresponds to the interaction energy between the individual adatoms given by (combinations of) Van-der-Waals, electrostatic (dipole), elastic, and electronic (substrate mediated) effects \cite{Nau05}. We safely can ignore this term, as STM experiments at $\sim$15 K have shown that the absolute value of the interaction energy drops below 0.1 meV for two Cu adatoms separated more than 60 {\AA} on a Cu(111) surface \cite{Rep00}. This is equivalent to an adatom density (fractional coverage) of $<$ 6.0 x 10$^{-4}$ ML, and as it will be shown in the following, we never reach such densities during the deposition.
\begin{figure}[h!tb]
\begin{center}
\epsfig{file=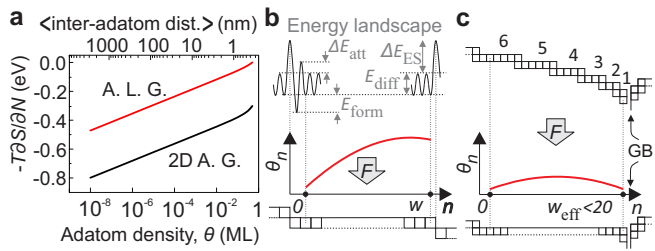,width=8.5714cm} \caption{{\bf Surface CP on terraces and next to GBs:} ({\bf a}) entropic component of the CP on Cu(111), estimated by the `adatom lattice gas' and the `2D adatom gas' model. ({\bf b}) Adatom density profile, ${{\theta }_{n}}$, on a terrace; the corresponding energy landscape is indicated at the top. ({\bf c}) Typical step configuration in the vicinity of a GB caused by the `Zeno effect' (top).  As $\Delta {{E}_{\mathrm{ES}}}$ vanishes for terrace widths with $w<6$, the combination of these terraces can be approximated with one effective terrace (bottom). This leads to a `lower' adatom density near the GB and, therefore, to `lower' stresses.}
\label{fig2}
\end{center}
\end{figure}

The third term, $-T\times \partial {{S}_{\mathrm{adatom}}}/\partial N$, involves the entropic effects of the 2D adatom gas. In general, depending on the adatom mobility, adatoms can be assumed to be confined on discreet lattice sites (`adatom lattice gas') or to be delocalized behaving as a 2D Van-der-Waals surface gas (`2D adatom gas') \cite{Iba06} (see Supplementary Note 1). As these two models naturally set the lower and upper limits for the adatom gas entropy, we calculated the boundary values of the CP for copper in Fig. \ref{fig2}a. Although the absolute values differ more than 0.3 eV, both models show a linear behavior in this logarithmic plot below 0.01 ML such that the following approximation holds for the CP variations:
\begin{equation}
\label{equ3}
\Delta \left[ -T\frac{\partial S}{\partial N} \right]\approx {{k}_{\mathrm{B}}}T\ln \left( {{\theta }_{\mathrm{2}}}/{{\theta }_{\mathrm{1}}} \right)
\end{equation}

\subsection{Adatom Density on Terraces}
To calculate the adatom densities during deposition and interruption, Fig. \ref{fig2}b shows a simplified model of the film surface, in which we define the position of the first lattice row next to the ascending step edge as the origin of a terrace with width $w$ in lattice units. By solving the differential equation for mass conservation, the adatom density at site $n$ on the terrace, ${{\theta }_{n}}$, can be derived as a function of deposition flux $F$ (see Supplementary Note 2):
\begin{equation}
\label{equ5}
{{\theta }_{n}}={{\theta }_{\mathrm{eq}}}+\frac{F\,w\,(a\,n+1)(s\,w+2)}{2{{\nu }_{\mathrm{d}}}\left( a\,s\,w+a+s \right)}-\frac{F{{n}^{2}}}{2{{\nu }_{\mathrm{d}}}}
\end{equation}
, where  ${{\theta }_{\mathrm{eq}}}=\exp \left( -{{E}_{\mathrm{form}}}/{{k}_{\mathrm{B}}}T \right)$, ${{\nu }_{\mathrm{d}}}={{\nu }_{\mathrm{0}}}\exp \left( -{{{E}_{\mathrm{diff}}}}/{{{k}_{\mathrm{B}}}T}\; \right)$, $a=\exp \left( -\Delta {{E}_{\mathrm{att}}}/{{k}_{\mathrm{B}}}T \right)$ and $s={{s}_{\mathrm{0}}}\exp \left( -\Delta {{E}_{\mathrm{ES}}}/{{k}_{\mathrm{B}}}T \right)$, in which ${{\nu }_{\mathrm{0}}}$, ${{E}_{\mathrm{diff}}}$, ${{E}_{\mathrm{form}}}$, $\Delta {{E}_{\mathrm{att}}}$, ${{s}_{\mathrm{0}}}$, $\Delta {{E}_{\mathrm{ES}}}$ are the diffusion rate prefactor, diffusion barrier, adatom formation energy from a kink site of the step, the attachment barrier, correction prefactor for hop over the step, and the Ehrlich-Schwoebel barrier (see Fig. \ref{fig2}b). Note that at zero deposition flux, the adatom density at each position $n$ of the terrace is equal to the equilibrium density ${{\theta }_{\mathrm{eq}}}$. For constant deposition with $\Delta E_{\mathrm{ES}} \gg \Delta E_{\mathrm{att}}$, $s\ll a$, the adatom density is highest close to the end of the terrace (see Fig. \ref{fig2}b), whereas the maximum shifts to the middle of the terrace for low values of $\Delta E_{\mathrm{ES}}$ (see Fig. \ref{fig2}c). Note that the maximum of the adatom density is only exactly at the end of the terrace for $E_{\mathrm{ES}} = \infty$.\ \\
Combining Eqs. \ref{equ2} to \ref{equ5}, one can calculate $\Delta {{\mu }_{\mathrm{s}}}$ as a function of deposition flux for any site n on the terrace.

\subsection{Chemical Potential of the Grains}
Considering an in-plane isotropic biaxial film ($\sigma _{\mathrm{x}}=\sigma_{\mathrm{y}}$ and $\sigma_{\mathrm{z}}=0$), it can be proven for the right hand side of Eq. \ref{equ1} that the CP within the grains is proportional to its total internal stress level $\sigma_{\mathrm{g}}$, (see Supplementary Note 3):
\begin{equation}
\label{equ7}
\Delta {{\mu }_{\mathrm{g}}}=-\int_{{\bf \sigma}_{\mathrm{rs}}^{\mathrm{0}}}^{{\bf \sigma}_{rs}^{\mathrm{0}}+ \Delta {\bf \sigma}_{rs}^{\mathrm{rev}}}{\bf \Omega}_{ij}d{\bf \sigma}_{ij}=-\Omega_{\mathrm{x}}\Delta{{\sigma }_{\mathrm{x}}}-\Omega_{\mathrm{y}}\Delta{{\sigma }_{\mathrm{y}}}\approx -\frac{2}{3}\Omega\Delta{{\sigma }_{\mathrm{g}}}
\end{equation}
, where $\Omega$ is the atomic volume.

\subsection{Derivation of the Stress Jumps}
Motivated by the fact that stress emergence has to be GB related \cite{Lei09}, it is crucial to derive the stress jumps using the $\Delta {{\mu }_{\mathrm{s}}}$  immediately next to the GBs! Combining the above equations at position $n=w$, the predicted upper limit for the reversible stress jumps based on pure thermodynamics is given by
\begin{equation}
\label{equ8}
\left| \Delta {{\sigma }_{\mathrm{rev}}} \right|\approx \frac{3}{2}\frac{k_{\mathrm{B}}T}{\Omega }\ln \left( \text{1}+\frac{F\,w\left( a\,w+2 \right)}{2{{\nu }_{\mathrm{d}}}\left( a\,s\,w+s+a \right){{\theta }_{\mathrm{eq}}}} \right)
\end{equation}
It has been shown for gold at room temperature that step-flow growth is the underlying atomic process for polycrystalline film growth in the VW regime \cite{Ros07}. On the basis of the so-called `Zeno effect', terraces closer to the GB get progressively decreased in their width during depostion, which leads to an enhanced surface curvature at the GB vicinity, as sketched in Fig. \ref{fig2}c-top  \cite{Ros07,Elk93}. This results in a deviation from the macroscopic equilibrium surface of an annealed polycrystalline film \cite{Ros03}. Although not mentioned in \cite{Ros07}, this deviation changes only slightly back over one hour upon stopping the deposition while keeping the film at room temperature. The same effect, which is due to the existence of a significant Ehrlich-Schwoebel barrier, has also been observed on Cu(111) \cite{Gie99-2}. For our study, we, therefore, safely omit Gibbs-Thompson correction terms associated with macroscopic surface curvature variations.

\begin{figure}[h!tb]
\begin{center}
\epsfig{file=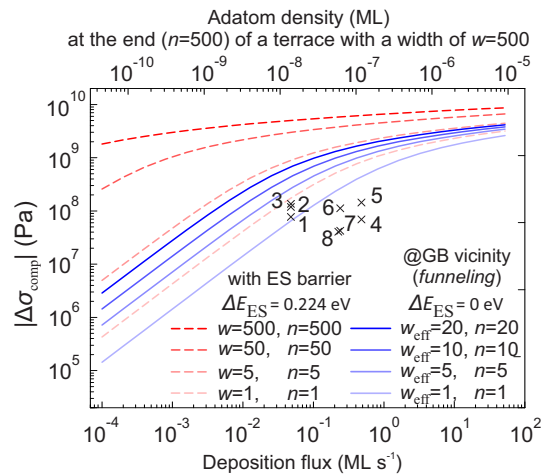,width=7.0088cm} \caption{{\bf Reversible stress jumps for a (111) textured copper film as a function of flux:}  The dashed red curves are calculated for typical terraces ($\text{1}<w<\text{5}00$ and $\Delta {{E}_{\mathrm{ES}}}=0.224\,\,\text{eV}$), whereas the solid blue curves describe terraces in the vicinity of GBs (${{w}_{\mathrm{eff}}}\le 20$ and $\Delta {{E}_{\mathrm{ES}}}=0\,\,\text{eV}$ ) where funneling takes place. The top horizontal axis shows the adatom density for a typical surface terrace ($w=500$ and $\Delta {{E}_{\mathrm{ES}}}=0.224\ \text{eV}$). Crosses show experimentally reported literature values: 1,2,3 \cite{Shu96}, 4 \cite{Abe85}, 5,6 \cite{Fri04}, 7 \cite{Cho05-2}, and 8 \cite{Glad06}. Note that the stresses are `lower', if it is `easier' for atoms to diffuse to and into GBs.}
\label{fig3}
\end{center}
\end{figure}

The red dashed lines in Fig. \ref{fig3} show the predicted stress jumps derived via Eq. \ref{equ8} for surface terrace widths between 1 and 500 atomic spacings and the existence of an Ehrlich-Schwoebel barrier. It is striking that the obtained stress values exceed even the experimentally ones (crosses) and that we receive a rather good agreement already for a terrace width with $w=1$ atomic spacing. To get a feeling for the numbers, the top horizontal axis shows the adatom density at the end point ($w = n$) for a terrace with a width of $w=500$ (74 nm). Note that the adatom density is less than $10^{-4}$ ML even for fluxes as high as 100 ML\ $\mathrm{s^{-1}}$. This validates our dilute adatom gas assumption while calculating the surface CP and justifies our steady-state thermodynamic approach. Note also that it is surprising that a dilute adatom gas of less than $10^{-4}$ ML has the potential to induce $\sim$1 GPa stress in the film, which is more than both the yield and the ultimate strength of copper! Please note that this stress can be realized in the bulk only, if there exists a kinetically not limited atomic mechanism that transfers the CP variation of the surface to the grain interior. The curves are calculated for Cu(111) and we have used T = 298 K, $\Omega = \frac{1}{4}(361.49 $ pm$)^{3}$, ${{\nu }_{\mathrm{0}}}={{10}^{12}}$ Hz, ${{E}_{\mathrm{diff}}}=0.0\text{4}0\,\text{eV}$ \cite{Kno02}, ${{s}_{\mathrm{0}}}=15$ \cite{Iba06}, $\Delta {{E}_{\mathrm{ES}}}=0.\text{224 eV}$ \cite{Gie99}, $\Delta {{E}_{\mathrm{att}}}=0\,\,\text{eV}$ ($\Delta E_{\mathrm{att}}\approx 0$ for most metals at room temperature), and ${{E}_{\mathrm{form}}}=0.\text{714}\,\,\text{eV}$ \cite{Sto94}, for our calculations, which implies an adatom equilibrium density of ${{\theta }_{\mathrm{eq}}}=8.6\times {{10}^{-13}}$ ML at zero deposition flux.

It is known that stress relaxation mechanisms are active both during the growth and after stopping the deposition, which reduce the absolute intrinsic film stress \cite{Cho05-2,Koc86}. Indeed, the experimental stress values (black crosses) are in general lower than the red dashed stress lines. However, if one considers that the time constant of the stress relaxation processes are distinctively larger than the time constant of the reversible jumps \cite{Yu14-2}, the discrepancy between the observed and the predicted values for typical surface terraces is too large to be explained by the stress relaxation effects alone.

\subsection{Stress Jumps considering Funneling}
As our derivation of the stress jumps exceeds in general the experimental values, we turn our attention to experiments on both Cu(111) and Ag(111) at room temperature: these experiments revealed that the Ehrlich-Schwoebel barrier of the lower step vanishes,if the distance between two neighboring steps becomes less than 6 atomic spacings \cite{Gie98,Gie00}! This effect opens a fast mass transfer channel for terraces with $w\leq 5$ called `funneling'. The fast mass transport region can extend up to 21 atomic sites (over 5 steps) away from the GB. We account for this by setting the critical terrace width to $w=5$ and by assuming that the Zeno effect reduces the width of the subsequent terraces by one atomic spacing. With this approximations, the last 6 terraces next to a GB can be treated as one single terrace with an `effective' width of $w_{\mathrm{eff}}\simeq 20$, with $\Delta {{E}_{\mathrm{ES}}}=0$ ($s$ = 1), see Fig. \ref{fig2}c bottom. However, as intermediate step edges given by the 5 steps can potentially act as adatom sinks, we evaluate the stress jumps for effective terrace widths, $w_{\mathrm{eff}}$, ranging 1 to 20 atomic spacings (see blue lines in Fig. \ref{fig3}). As our calculations define upper limits for the stress jumps, a funneling width of 5 spacings represents the best fit. This means that is it enough, if only the last 2 steps (and not 5) before the GB show funneling. The fact that our result with the inclusion of funneling delivers a rather good fit with the experimental values is a strong indication for the validity of the GB insertion model, especially, as we calculate the CP on the surface exactly next to the GBs. Note, however, that we do not address the exact atomic pathway (e.g. diffusion, exchange,...), as we evaluate only the thermodynamics.

\section{Discussion}
Although we `lowered the barriers' for the atoms to diffuse towards and into the GB, we receive `lower stresses' than without funneling. This seemingly counterintuitive behavior demonstrates that we are not addressing a certain atomic diffusion/incorporation model, but calculate the thermodynamic driving force on the basis of the CP. Lowering the barriers for atoms to diffuse towards and into the GB, decreases the adatom density near the GB and results in a reduction of the driving force for the atoms to diffuse into the GB. Evidently, the funneling curves predict the correct order of magnitude for the stress levels. By setting $(s=a=1)$ in Eq. \ref{equ5}, one gets an estimate of the adatom density at the GB vicinity $n=w_{\mathrm{eff}}$: for a deposition rate of 1 ML\ $\mathrm{s^{-1}}$ the adatom density is predicted to be lower than $10^{-10}$ ML!

Finally we address another hypothetical mechanism in combination with the equilibrium situation between the surface and the bulk. For materials with a low enough Ehrlich-Schwoebel barrier and funneling terraces, the CP distribution shows a maximum around the middle of the terrace, see Fig. 2c. In addition, the CP at the end position of, e.g., a large terrace in the middle of the grain (with Ehrlich-Schwoebel barrier) is significantly larger than the CPs of funneling terraces that connect to the grain boundaries. Pure thermodynamic considerations imply that also these maxima tend to establish equilibrium with the interior of the grain underneath. As the grain is comparable to a single crystal, dislocation nucleation would be an imaginable pathway to balance the CP differences. However, on epitaxial films and single crystals the reversible stress jumps are not observed \cite{Lei09}. The reason for this is a high nucleation barrier: a critical stress of 1.3 GPa has been determined to nucleate dislocations in Cu at 300 K \cite{Zhu08}. Such high absolute stress values are neither observed experimentally, nor does our model predict an equivalent rise of the surface CP (except for large terraces in combination with high deposition fluxes). Dislocation nucleation is, therefore, kinetically limited! The overall picture is that during the growth all points on the surface are in parallel trying to balance their CPs with adjacent positions as well as with the grain underneath. However, in which way and by which rate the CP of the grains will change clearly depends on the rates of the pathways between all subsystems: surface, GB, and bulk. With a significant dislocation nucleation barrier and active GB diffusion as well as atom incorporation, the whole system quickly evolves towards equilibrium via atom insertion in GBs. As a result the CP difference between the grain and the surface CP maxima are reduced, which effectively lowers the driving force for dislocation nucleation making this latter process even less favorable.

If one intends to compare our results with experiments, it is important to realize that we determine only the pure reversible equilibrium jumps. For a proper comparison the experiments should have no kinetic limitation of atoms going in/out of the grain boundaries, should be performed long enough such that equilibrium has reached (all GBs show the equilibrium density of additional atoms), and no stress relaxation mechanisms should occur. In this limit, we expect the stress jumps to be GB density independent. Kinetic limitations would immediately result in a GB density dependence, as equilibrium will not be reached and the rate towards equilibrium scales with the number of the pathways and hence the GB density.\\ The deposition flux and temperature dependence is more complex, as the growth mode (layer-by-layer, step flow, 3D/rough growth) that determines the size of the terrace next to the GBs, also changes with deposition rate and mobility. If one, e.g., lowers the rate for a film that growths in 3D growth mode, one might enter step flow growth conditions in which effective larger terraces (with higher adatom density) might communicate with the GBs such that the stress jumps are even higher instead of lower.

Our analysis shows that entropic effects in the extremely dilute adatom gas on the surface of a polycrystalline film during vapor deposition are strong enough to cause plastic deformation in the film. The predicted film stresses are even higher than the observed ones. If we `lower the barriers' for atoms to diffuse towards and into the GB by funneling, the `stresses decrease' and the predicted values perfectly match the experimental ones. With this we deliver the, until know missing thermodynamic driving force for any GB atom insertion model. Further experimental research, similar to \cite{Ros07,Yu14}, is needed to clarify the exact atomistic mechanisms and pathways behind this effect.

\section{Acknowledgements}
We acknowledge R.V. Mom for discussions about thermodynamics and chemical potentials. The research described in this paper has been solely performed at the University of Leiden within the "vidi" project of M.J.R. that was financed via by the Dutch Technology Foundation STW (Project No. 10779), which is the applied science division of NWO, and the Technology Program of the Ministry of Economic Affairs.
\section{Author contributions}
The project was initiated and conceptualized by M.J.R. In his discussions with A.S., he pointed out the deficiencies of stress generation models at the time. Inspired by the discussions, A.S. envisioned a possible path to calculate the reversible stress analytically. Encouraged by this, the authors developed the thermodynamical description together, for which A.S. worked out the equations. Together they discussed the approach, interpreted the results, and wrote the manuscript.
\section{Additional information}
{\bf Supplementary Information} accompanies this paper at http://www.nature.com/naturecommunications\\
{\bf Competing financial interests:} The authors declare no competing financial interest.\\
{\bf Reprints and permission} information is available online at http://npg.nature.com/reprintsandpermissions/ \\
{\bf How to cite this article:} XXXXX \\


\begin{references}
\bibitem{Flo02} Floro, J. A., Chason, E., Cammarata, R. C. \& Srolovitz, D. J. Physical origins of intrinsic stresses in Volmer–Weber thin films. {\it MRS Bulletin} {\bf 27,} 19-25 (2002).
\bibitem{Spa00} Spaepen, F. Interfaces and stresses in thin films. {\it Acta Mater.} {\bf 48,} 31-42 (2000).
\bibitem{Shu96} Shull, A. L. \& Spaepen, F. Measurements of stress during vapor deposition of copper and silver thin films and multilayers. {\it J. Appl. Phys.} {\bf 80,} 6243-6256 (1996).
\bibitem{Lau81} Laugier, M. Intrinsic stress in thin films of vacuum evaporated LiF and ZnS using an improved cantilevered plate technique. {\it Vacuum} {\bf 31,} 155-157 (1981).
\bibitem{Cam00} Cammarata, R. C., Trimble, T. M. \& Srolovitz, D. J. Surface stress model for intrinsic stresses in thin films. {\it J. Mater. Res.} {\bf 15,} 2468-2474 (2000).
\bibitem{Hof76} Hoffman, R. W. Stresses in thin films: the relevance of grain boundaries and impurities. {\it Thin Solid Films} {\bf 34,} 185-190 (1976).
\bibitem{Fre01} Freund L. B. \& Chason, E. Model for stress generated upon contact of neighboring islands on the surface of a substrate. {\it J. Appl. Phys.} {\bf 89,} 4866-4873 (2001).
\bibitem{Abe90} Abermann, R. Measurements of the intrinsic stress in thin metal films. {\it Vacuum} {\bf 41,} 1279-1282 (1990).
\bibitem{Fri04} Friesen, C., Seel, S. C. \& Thompson, C. V. Reversible stress changes at all stages of Volmer–Weber film growth. {\it J. Appl. Phys.} {\bf 95,} 1011-1020 (2004).
\bibitem{abe79} Abermann, R., Koch, R. \& Kramer, R. Electron microscope structure and internal stress in thin silver and gold films deposited onto MgF2 and SiO substrates. {\it Thin Solid Films} {\bf 58,} 365-370 (1979).
\bibitem{Cha02} Chason, E., Sheldon, B. W., Freund, L. B., Floro, J. A. \& Hearne, S. J. Origin of compressive residual stress in polycrystalline thin films. {\it Phys. Rev. Lett.} {\bf 88,} 156103 (2002).    
\bibitem{Tel07} Tello, J. S., Bower, A. F., Chason, E. \& Sheldon, W. Kinetic Model of Stress Evolution during Coalescence and Growth of Polycrystalline Thin Films. {\it Phys. Rev. Lett.} {\bf 98,} 216104 (2007).
\bibitem{Pao07} Pao, C.-W., Foiles, S. M., Webb III, E. B., Srolovitz, D. J. \& Floro, J. A. Thin Film Compressive Stresses due to Adatom Insertion into Grain Boundaries. {\it Phys. Rev. Lett.} {\bf 99,} 036102 (2007).    
\bibitem{Gon13} Gonz\'{a}lez-Gonz\'{a}lez, A. et al. E. Postcoalescence evolution of growth stress in polycrystalline films. {\it Phys. Rev. Lett.} {\bf 110,} 056101 (2013).
\bibitem{Yu14} Yu, H. Z. \& Thompson, C. V. Correlation of shape changes of grain surfaces and reversible stress evolution during interruptions of polycrystalline film growth. {\it Appl. Phys. Lett.} {\bf 104,} 141913 (2014).
\bibitem{Lei09} Leib, J., M\"{o}nig, R. \& Thompson, C. V. Direct evidence for effects of grain structure on reversible compressive deposition stresses in polycrystalline gold films. {\it Phys. Rev. Lett.} {\bf 102,} 256101 (2009).
\bibitem{Lei10} Leib, J. \& Thompson, C. V. Weak temperature dependence of stress relaxation in as-deposited polycrystalline gold films. {\it Phys. Rev. B} {\bf 82,} 121402(R) (2010).
\bibitem{Pao06} Pao, C. W., Srolovitz, D. J. \& Thompson, C. V. Effects of surface defects on surface stress of Cu(001) and Cu(111). {\it Phys. Rev. B} {\bf 74,} 155437 (2006).
\bibitem{Sek04} Sekerka, R. F. {\it Crystal Growth - From Fundamentals to Technology.} (editors, M{\"u}ller, G., M{\`e}tois, J. J. \&  Rudolph, P.) Ch.1 (Elsevier Science, 2004).
\bibitem{Gie99-2} Giesen, M. \& Ibach, H. Step edge barrier controlled decay of multilayer islands on Cu(111). {\it Surf. Sci.} {\bf 431,} 109-115 (1999).
\bibitem{Zha90} Zhang, J. \& Nancollas, G. H. Kink densities along a crystal surface step at low temperatures and under nonequilibrium conditions. {\it J. Cryst. Growth} {\bf 106,} 181-190 (1990).
\bibitem{Ros07} Rost, M. J. In situ real-time observation of thin film deposition: roughening, Zeno effect, grain boundary crossing barrier, and steering. {\it Phys. Rev. Lett.} {\bf 99,} 266101 (2007).
\bibitem{Mic04} Michely, T. \& Krug, J. {\it Islands, Mounds and Atoms.} Ch. 4.3 (Springer, 2004).
\bibitem{Abe80} Abermann, R., Martinz, H. P. \& Kramer, R. Thermal effects during the deposition of thin silver, gold and copper films and their influence on internal stress measurements. {\it Thin Solid Films} {\bf 70,} 127-137 (1980).
\bibitem{Nau05} Naumovets, A. G. Collective surface diffusion: an experimentalist’s view. {\it Physica A} {\bf 357,} 189-215 (2005).
\bibitem{Rep00} Repp, J. et al. Substrate Mediated Long-Range Oscillatory Interaction between Adatoms: Cu(111). {\it Phys. Rev. Lett.} {\bf 85,} 2981-2984 (2000).
\bibitem{Iba06} Ibach, H. {\it Physics of Surfaces and Interfaces.} Chs. 4.3, 5.4, 10.1, 10.4, and 11.4 (Springer 2006)
\bibitem{Ros03} Rost, M. J., Quist, D. \& Frenken, J. W. M. Grains, growth, and grooving. {\it Phys. Rev. Lett.} {\bf 91,} 026101 (2003).
\bibitem{Elk93} Elkinani, I. \& Villain, J. Le paradoxe de Zenon d'Elee {\it Solid State Comm.} {\bf 87,} 105-108 (1993).
\bibitem{Gie98} Giesen, M., Schulze Icking-Konert, G. \& Ibach, H. Fast decay of adatom islands and mounds on Cu(111): a new effective channel for interlayer mass transport. {\it Phys. Rev. Lett.} {\bf 80,} 552-555 (1998).
\bibitem{Gie00} Giesen, M. \& Ibach, H. On the mechanism of rapid mound decay. {\it Surf. Sci.} {\bf 464,} L697-L702 (2000).
\bibitem{Abe85} Abermann, R. \& Koch, R. The internal stress in thin silver, copper and gold films. {\it Thin Solid Films} {\bf 129,} 71-78 (1985).
\bibitem{Glad06} Gladyszewski, G., Chocyk, D., Proszynski, A. \& Pienkos, T. {\it Microelec. Eng.} {\bf 83,} 2351-2354 (2006).
\bibitem{Kno02} Knorr, N. et al. Long-range adsorbate interactions mediated by a two-dimensional electron gas {\it K. Phys. Rev. B} {\bf 65,} 115420 (2002).
\bibitem{Gie99} Giesen, M., Schulze Icking-Konert, G. \& Ibach, H. Interlayer mass transport and quantum confinement of electronic states. {\it Phys. Rev. Lett.} {\bf 82,} 3101-3104 (1999).
\bibitem{Sto94} Stoltze, P. Simulation of surface defects. {\it J. Phys.: Condens. Matter} {\bf 6,} 9495-9517 (1994).
\bibitem{Cho05-2} Chocyk, D., Proszynski, A., Gladyszewski, G., Pienkos, T. \& Gladyszewski, L. Post-deposition stress evolution in Cu and Ag thin films. {\it Optica Applicata} {\bf XXXV,} 419-424 (2005).
\bibitem{Koc86} Koch, R. \& Abermann, R. Microstructural changes in vapour-deposited silver, copper and gold films investigated by internal stress measurements. {\it Thin Solid Films} {\bf 140,} 217-226 (1986).
\bibitem{Tho96} Thompson, C. V. \& Carel, R. Stress and grain growth in thin films. {\it J. Mech. Phys. Solids} {\bf 44,} 657-673 (1996).
\bibitem{Yu14-2} Yu, H. Z., Leib, J. S., Boles, S. T. \& Thompson, C. V. Fast and slow stress evolution mechanisms during interruptions of Volmer-Weber growth. {\it J. Appl. Phys.} {\bf 115,} 043521 (2014).
\bibitem{Kur01} K\"urpick, U. Self-diffusion on (100), (110), and (111) surfaces of Ni and Cu: A detailed study of prefactors and activation energies. {\it Phys. Rev. B} {\bf 64,} 075418 (2001).
\bibitem{Zhu08} Zhu, T., Li, J., Samanta, A., Leach, A., \& Gall, K. Temperature and strain-rate dependence of surface dislocation nucleation. {\it Phys. Rev. Lett.} {\bf 100,} 025502 (2008).
\bibitem{San12} Sandstr\"{o}m R. \& Hallgren, J. The role of creep in stress strain curves for copper. {\it J. Nuc. Mater.} {\bf 422,} 51-57 (2012).
\bibitem{Eli08} Eliaz, N. \& Banks-Sills, L. Chemical potential, diffusion and stress - common confusions in nomenclature and units. {\it Corros. Rev.} {\bf 26,} 87-103 (2008).

\end{references}
\end{document}